\documentclass[11pt]{article}
\usepackage{amssymb}
\usepackage{multicol}
\usepackage{amsmath}
\usepackage{longtable}

\newtheorem{example}{Example}[section]

\begin{document}
\begin{center}
{\LARGE\bf Accurate inference for a one parameter distribution}\\[1ex]
{\LARGE\bf based on the mean of a transformed sample}\\[1ex]
by\\[1ex]
Christopher S. Withers\\
Applied Mathematics Group\\
Industrial Research Limited\\
Lower Hutt, NEW ZEALAND\\[2ex]
Saralees Nadarajah\\
School of Mathematics\\
University of Manchester\\
Manchester M13 9PL, UK
\end{center}
\vspace{1.5cm}
{\bf Abstract:}~~A great deal of inference in statistics is based on making the
approximation that a statistic is normally distributed.
The error in doing so is generally $O(n^{-1/2})$ and can be very
considerable when the distribution is heavily biased or skew.
This note shows how one may reduce this error to $O(n^{-(j+1)/2})$,
where $j$ is a given integer.
The case considered is when the
statistic is the mean of the sample values from a continuous one-parameter
distribution, after the sample has undergone an initial transformation.


\section{Introduction and summary}
\setcounter{equation}{0}

Given a random sample of size $n$, the usual confidence interval for the population mean is of the order $O(n^{-1/2})$.
The aim of this note is to show how a more accurate confidence interval of the order $O(n^{-(j+1)/2})$,
where $j$ is a given integer, can be obtained.
This is an important problem because finding accurate confidence intervals for
the population mean is an everyday problem faced by many scientists and engineers.

Suppose for some statistic $\psi_n$,
$Y_n (\theta) = n^{1/2} \{ \psi_n - g(\theta) \} / \sigma (\theta) \rightarrow N(0,1)$
as  $n \rightarrow \infty$, where $\theta \in R$.
Cornish and Fisher (1937) and Fisher and Cornish (1960) obtained an Edgeworth type expansion for
the distribution of $Y_n (\theta)$, and an asymptotic expansion for its
percentile points when this distribution is parameter-free.
Withers (1984) gave a simplified version of their results which
reduced the labor of their application.
Withers (1983) considered the more general case, where the distribution of $Y_n (\theta)$
does depend on $\theta$.
In this case a parameter-free transformation $V_{nx} (\cdot)$ was given such that
\begin{eqnarray}
V_{nx_1} \left( \psi_n \right) < \theta < V_{nx_2} \left( \psi_n \right)
\mbox{ with probability } 1 - \alpha
\label{1.1}
\end{eqnarray}
provided
\begin{eqnarray}
\Phi \left( x_1 \right) - \Phi \left( x_2 \right) = 1 - \alpha, \mbox{ e.g. } x_1 = -x_2 = \Phi^{-1} \left( 1-\alpha/2 \right),
\label{1.2}
\end{eqnarray}
assuming $g(\cdot)$ is an increasing function, where $\Phi (\cdot)$ denotes the distribution of a standard normal random variable.
If $g(\cdot)$ is decreasing the inequalities in (\ref{1.1}) are reversed.

In Section 2, we show how this theory applies to $\psi_n = \overline{X}_n$,
the mean of $X_1, \ldots, X_n$ i.i.d. $F(x,\theta)$ on $R$, where $F(x,\theta)$ is of known parametric form.

In applications $\{ X_i \}$ will generally not be the original
observations $\{ Y_i \}$, say, but will be given by $X_i = h(Y_i)$, where $h(\cdot)$ is a transformation chosen from considerations of
efficiency, robustness or ease of computation of the first few
cumulants of $F(x,\theta)$ as functions of $\theta$.
So, if $Y_1 \sim R(x,\theta)$ then
\begin{eqnarray}
F(x,\theta) = R \left( h^{-1} (x), \theta \right)
\label{1.3}
\end{eqnarray}
for $h(\cdot)$ one to one increasing.

However, $\{ Y_i \}$ need not lie in $R$.
Their distribution may, in fact, depend on parameters, $\lambda$, other than
$\theta$ provided $F(x, \theta)$ does not depend on $\lambda$.
Note that $\theta$ itself may be a reparameterisation of an original parameter.

In Section 3 the efficiency and robustness of this class of procedures is considered.

In Section 4 this theory is applied to the `Lehmann alternative':
$F(x,\theta) = R(x)^\theta$, where by suitable choice of $h(\cdot)$, $R(\cdot)$ may be any continuous distribution.

For many parameter inference problems see Withers (1989).

\section{The general case}
\setcounter{equation}{0}

Let $X_1, \ldots, X_n$ be a random sample for a distribution $F(x, \theta)$
on $R$ such that $g(\theta) = EX_1$ is a known one-to-one function from the parameter space, assumed to be some subset of $R$.
Set
\begin{eqnarray*}
\sigma(\theta)^2 = \mbox{var} X_1,
\
T_n = \overline{X}_n - g(\theta),
\
Y_n (\theta) = n^{1/2} T_n / \sigma (\theta).
\end{eqnarray*}
For any real random variable $X$ set
\begin{eqnarray}
&&
K_r (X) = r \mbox{th cumulant of } X,
\nonumber
\\
&&
\ell_r (X) = \kappa_2 (X)^{-r/2} \kappa_r (X) - \delta_{r,2},
\label{2.1}
\end{eqnarray}
where $\delta_{r,s} = 1$ if $r = s$ and $\delta_{r, s} = 0$ if $r \neq s$.
Suppose that for some $j \geq 0$, $\kappa_r (X_1)$ exists for $1 \leq r \leq j+2$ and
\begin{eqnarray}
\raisebox{-1mm}{${\lim \sup} \atop {t \rightarrow \pm \infty}$} \left|
\int \exp (itx) dF (x, \theta) \right| <1.
\label{2.2}
\end{eqnarray}
This condition rules out many discrete lattice distributions.
Then by Theorem 3, page 541 of Feller (1971),
\begin{eqnarray*}
P \left( Y_n (\theta) \leq x \right) = \Phi (x) - \phi (x) \sum_{r=1}^j n^{-r/2} U_r (x) + o \left( n^{-j/2} \right)
\mbox{ as } n \rightarrow \infty
\end{eqnarray*}
uniformly in $x$, where $U_r (x)$ is a polynomial in $x$ defined in
terms of $\ell_1, \ldots, \ell_{r+2}$, where $\ell_r = \ell_r (X_1)$.
Note that $\ell_1 = \ell_2 = 0$.
($U_r = R_{r+2}$ is defined by Feller).
In particular, by equation (6.50) of Stuart and Ord (1987),
\begin{eqnarray*}
&&
U_1 = \ell_3 H_2 / 6,
\\
&&
U_2 = \ell_4 H_3 / 24 + \ell_3^2 H_5 /72,
\\
&&
U_3 = \ell_5 H_4 / 120 + \ell_3 \ell_4 H_6 / 144 + \ell_3^3 H_8 / 1296,
\\
&&
U_4 = \ell_6 H_5 / 720 + \ell_4^2 H_7 / 1152 + \ell_3 \ell_5 H_7 / 720 + \ell_3^2 \ell_4 H_9 / 1728 + \ell_3^4 H_{11} / 31104,
\end{eqnarray*}
where $H_r$ is the Hermite polynomial:
$H_r (x) = \exp (x^2/2) (-\partial / \partial x)^r \exp (-x^2/2)$.
For example, $H_1, \ldots, H_{10}$ are given by equation (6.23) of Stuart and Ord (1987).
Cornish and Fisher (1937) used this to show (for a more general
situation but assuming all cumulants exist) that
\begin{eqnarray*}
P \left( Y_n (\theta) \leq x \right) \equiv \Phi \left( \xi_n (x) \right) = \Phi \left( \xi_{nj} (x) \right) + o \left( n^{-j/2} \right),
\end{eqnarray*}
where
\begin{eqnarray*}
\xi_n (x) = x-\sum^\infty_{r=1} n^{-r/2} f_r (x),
\
\xi_{nj} (x) = x-\sum^j_{r=1} n^{-r/2} f_r (x)
\end{eqnarray*}
and $f_r (x)$ is a polynomial in $x$ depending on $\ell_1, \ldots, \ell_{r+2}$:
\begin{eqnarray*}
f_r (x) = \beta_r' a_r (x),
\end{eqnarray*}
where
\begin{eqnarray}
&&
\beta_1' = \ell_3,
\
\beta_2' = \left( \ell_4, - \ell_3^2 \right),
\
\beta_3' = \left( \ell_5, - \ell_3 \ell_4, \ell_3^3 \right),
\nonumber
\\
&&
\beta_4' = \left( \ell_6, -\ell_3 \ell_5, \ell_3^2 \ell_4, -\ell_3^4 \right)
\label{2.11}
\end{eqnarray}
and
\begin{eqnarray*}
&&
a_1 (x) = H_2/6,
\
a_2 (x) = \Big( H_3 (x)/24, \left( 4x^3 - 7x \right)/36 \Big)',
\\
&&
a_3 (x) = \Big( H_4 (x)/120, \left( 11x^4 -42x^2 +15 \right)/144, \left( 69x^4 - 187x^2 + 52 \right) / 648 \Big)',
\\
&&
a_4 (x) = \Big( H_5 (x)/720, \left( 5x^5 - 32x^3 +35x \right)/384, \left( 7x^5 - 48x^3 +51x \right)/360,
\\
&&
\qquad \qquad
\left( 111x^5 - 547x^3 + 456x \right)/864,
\left( 948x^5 - 3628x^3 + 2473x \right)/7776 \Big)'.
\end{eqnarray*}
They further showed that if $z_\alpha = \Phi^{-1} (1-\alpha)$ then
\begin{eqnarray*}
Y_n (\theta) \leq \eta_n \left( z_\alpha \right) \mbox{ with probability } 1-\alpha,
\end{eqnarray*}
where
\begin{eqnarray*}
\eta_n (y) = y+\sum_{r=1}^\infty n^{-r/2} g_r (y)
\end{eqnarray*}
and $g_r (y)$ is a polynomial in $x$ depending on $\ell_1, \ldots, \ell_{r+2}$:
\begin{eqnarray}
g_r (y) = \beta_r' b_r (y),
\label{2.16}
\end{eqnarray}
where
\begin{eqnarray*}
&&
b_1 (y) = H_2 (y)/6,
\\
&&
b_2 (y) = \Big( H_3 (y)/24, \left( 2y^3 - 5y \right)/36 \Big)',
\\
&&
b_3 (y) = \Big( H_4 (y)/120, \left( y^4 - 5y^2 + 2 \right)/24, \left( 12y^4 -53y^2 +17 \right)/324 \Big)',
\\
&&
b_4 (y) = \Big( H_5 (y)/720, \left( 3y^5 - 24y^3 + 29y z \right)/384, \left( 2y^5 - 17y^3 + 21y \right)/180,
\\
&&
\qquad \qquad
\left( 14y^5 - 103y^3 +107y \right)/288, \left( 252y^5 - 1688y^3 + 1511y \right)/7776 \Big)',
\nonumber
\end{eqnarray*}
and $g_5$, $g_6$ may be obtained from V, VI, pages 214, 215 of Fisher and Cornish (1960),
by setting $a=b=0$, $c=\ell_3$, $d=\ell_4$, $e=\ell_5$, etc.

So, under (\ref{2.2}), if
\begin{eqnarray}
\sigma = \sigma (\theta), \kappa_3 \left( X_1 \right), \ldots, \kappa_{j+2} \left( X_1 \right)
\mbox{ exist and do not depend on } \theta
\label{2.17}
\end{eqnarray}
then
\begin{eqnarray*}
G_n^{-1} (1-\alpha) = G_{nj}^{-1} (1-\alpha) + o \left( n^{-(j+1)/2} \right),
\end{eqnarray*}
where
\begin{eqnarray*}
G_n (x) = P \left( \overline{X}_n - g(\theta) \leq x \right),
\
G_{nj}^{-1} (1-\alpha) = n^{-1/2} \sigma \eta_{nj} \left( z_\alpha \right),
\
\eta_{nj} (y) = y+\sum_{r=1}^j n^{-r/2} g_r (y).
\end{eqnarray*}
In particular, the confidence interval
\begin{eqnarray}
g^{-1} \left( \overline{X}_n - G_{nj}^{-1} (\alpha / 2) \right) \leq \theta \leq g^{-1}
\left(\overline{X}_n - G_{nj}^{-1} (1-\alpha / 2) \right)
\label{2.19}
\end{eqnarray}
has level $1-\alpha + o(n^{-j/2})$, in fact $1-\alpha + O(n^{-(j+1)/2})$ if $\kappa_{j+3} (X_1)$ is finite.
More generally, if (\ref{2.17}) is weakened to allow $\sigma (\theta)$ to vary
with $\theta$ and $Y_n (\cdot)$ is one to one increasing (or
decreasing), and $x_1$, $x_2$ satisfy (\ref{1.2}), then a confidence
interval with level $1-\alpha+O(n^{-(j+1)/2})$ is
\begin{eqnarray}
Y_n^{-1} \left( \eta_{nj} \left( x_2 \right) \right) \leq \theta \leq Y_n^{-1} \left( \eta_{nj} \left( x_1 \right) \right)
\label{2.20}
\end{eqnarray}
with the inequalities reversed if $Y_n (\cdot)$ is decreasing.

These formulae have been shown to be extremely accurate.
One can judge the number of significant places when approximating $G_n^{-1}$
by $G_{nj}^{-1}$ by the size of the successive terms $n^{-r/2} g_r (z_\alpha)$, which generally alternate in sign.
See, for example, Fisher and Cornish (1960).

Withers (1983) -- for a more general situation -- showed how to
obtain a confidence interval for $\theta$ in the more
usual situations, where the cumulants depend on $\theta$.
This dependency is expressed by writing $g_r (x) = g_r (x, \theta)$, etc.

The main purpose of the present note is to apply these results to
the case of the sample mean under the assumptions (\ref{2.1}), (\ref{2.2}).
When the initial transformation $h(\cdot)$ is independent of $\theta$ then
applying Withers (1983) to $\overline{X}_n$, one obtains that a confidence
interval of level $1-\alpha+o(n^{-j/2})$ is
\begin{eqnarray}
V_{nx_1 j} \left( \overline{X}_n \right) < \theta < V_{nx_2 j} \left( \overline{X}_n \right),
\label{2.22}
\end{eqnarray}
where $V_{nxj} (t) = g^{-1} (S_{nxj} (t))$, $x_1$, $x_2$ satisfy (\ref{1.2}), and
\begin{eqnarray}
S_{nxj} (t) = t+\sum_{i=1}^{j+1} n^{-i/2} Q_i (t)
\label{2.23}
\end{eqnarray}
for $\{ Q_i (t) \}$ given by Withers (1983) in terms of $P_i (t) = \sigma (g^{-1} (t)) g_{i-1} (x, g^{-1} (t))$ for $\{ g_i \}$ as above,
where $g_0 (x, \theta) = x$.
Here, we have assumed $g(\cdot)$ to be increasing.
If $g(\cdot)$ is decreasing the inequalities in (\ref{2.22}) are reversed.

In particular, $Q_1 = -P_1$, $Q_2 = -P_2 -\dot{P}_1 Q_1$ and $Q_3 = -P_3 -\dot{P}_2 Q_1 - \ddot{P}_1 Q_1^2 /2$.

For such calculations it is convenient to write $P_i$ in the form
\begin{eqnarray*}
P_i (t) = M_i (t)' b_{i-1} (x),
\
i \geq 1,
\end{eqnarray*}
where $M_i (t) = m_i (g^{-1} (t))$, $m_i (\theta) = \sigma (\theta) \beta_{i-1}$, $\beta_0 = 1$ and $b_0 (x) =x$
with $\{ \beta_r, b_r \}$ given by (\ref{2.11}), (\ref{2.16}).
Setting $K_r (\theta) = \kappa_r (X_1)$, one obtains
\begin{eqnarray*}
&&
m_1 (\theta) = \sigma (\theta),
\
m_2 (\theta) = \sigma (\theta)^{-2} K_3 (\theta),
\\
&&
m_3 (\theta)' = \left( \sigma (\theta)^{-3} K_4 (\theta), \sigma (\theta)^{-5} K_3 (\theta)^2 \right),
\end{eqnarray*}
etc, and so
\begin{eqnarray}
&&
Q_1 (t) = -M_1 (t) x,
\label{2.29}
\\
&&
Q_2 (t) = -M_2 (t) b_1 (x) + x^2 D_t M_1 (t)^2 /2,
\label{2.30}
\\
&&
Q_3 (t) = -M_3 (t)' b_2 (x) + xb_1 (x) D_t M_1 (t) -x^3 D_t M_1 (t)^3 /6,
\label{2.31}
\end{eqnarray}
and so forth, where $D_t = \partial / \partial t$.

\section{Efficiency and robustness}
\setcounter{equation}{0}

So far our concern has been to obtain accurate inference on the
parameter of the original distribution $R(x,\theta)$ from the size of
$\overline{X}_n$, where $X_i \equiv h(Y_i)$ and $h(\cdot)$ is a given
transformation.
We now consider the efficiency of the procedure, and its robustness to outliers,
as these factors are important in the choice of $h(\cdot)$.

Let $F_n$ be the empirical distribution of $\{ Y_i \}$, which we
shall suppose lie in $R^s$.
Corresponding to (\ref{2.22}) is the point estimate
\begin{eqnarray*}
\widehat{\theta}_n = \theta \left( F_n \right),
\end{eqnarray*}
where $\theta (F) = g^{-1} (\int h \, df)$ and $g(\cdot)$ is fixed by the choice of $h(\cdot)$:
\begin{eqnarray*}
g(\theta) = \int x \, dF (x,\theta) = \int h(y) \, dR (y,\theta).
\end{eqnarray*}
The influence function of $\widehat{\theta}_n$ is
\begin{eqnarray*}
I_\theta (x,F) = \left( h(x) - \int h \, dF \right) / \dot{g} \left( g^{-1} \left( \int h \, dF \right) \right)
\end{eqnarray*}
which evaluated at $F(\cdot) = R(\cdot, \theta)$ gives
$I_\theta (x) = ( h(x) - g(\theta)) / \dot{g} (\theta)$.
So, to reduce the effect of outliers it is desirable that $h(\cdot)$ be bounded.
Also
\begin{eqnarray*}
n^{1/2} \left( \widehat{\theta}_n - \theta \right) \stackrel{\cal L}{\rightarrow}
{\cal N} \left( 0, V (\theta, h) \right)
\end{eqnarray*}
as $n \rightarrow \infty$, where
\begin{eqnarray}
V(\theta, h) = \int I_\theta (x)^2 \, dR (x, \theta) = \left( \int h (x)^2 \, dR (x, \theta) - g(\theta)^2 \right) / \dot{g} (\theta)^2 =
\sigma (\theta)^2 / \dot{g} (\theta)^2.
\label{3.5}
\end{eqnarray}

The asymptotic efficiency of $\widehat{\theta}_n$ or of the confidence
interval (\ref{2.19}) is inversely proportional to $V(\theta, h)$.
Note that $V(\theta, h)$ is minimized by $h=q_\theta$, where $q_\theta$ is Fisher's score function
\begin{eqnarray*}
q_\theta (x) = \partial / \partial \theta \log \, dR (x, \theta) / dR (x, 0).
\end{eqnarray*}

The maximum likelihood estimate $\theta_n^*$ is asymptotically
equivalent to this choice in the sense that $n^{1/2} (\theta_n^* - \widehat{\theta}_n) \stackrel{p}{\rightarrow} 0$ as $n \rightarrow \infty$.
However, the results of Section 2 have assumed $h(\cdot)$ is independent of $\theta$, so can only be applied to
$\theta_n^*$ when $q_\theta (x)$ has the form $a(\theta) b(x)$.

\section{Lehmann's alternative}
\setcounter{equation}{0}

In this section, we illustrate the results of Sections 2 and 3 when the
original sample $\{ Y_i \}$ has distribution $R(x,\theta) = F_0 (x)^\theta$,
where $\theta >0$, and $F_0 (\cdot)$ is a continuous distribution.
This is sometimes known as `Lehmann's alternative'.
By (\ref{1.3}), $\{X_i = h(Y_i) \}$ have distribution $F(x,\theta) = R(x)^\theta$,
where $R(x) = F_0 (h^{-1} (x))$.
So, by suitable choice of $h(\cdot)$, $R(\cdot)$ may be chosen to be any continuous
distribution on $R$.
The cumulative generating function for $X_1$ is
\begin{eqnarray*}
K_R (t) = \log \int \exp (tx) \, dR(x)^\theta.
\end{eqnarray*}
However, it is sometimes easier to calculate the necessary cumulants directly.

The maximum likelihood estimate is given by
$\theta_n^* = -\overline{X}_n^{-1}$, where $h(x) = \log F_0 (x)$.
This yields

\begin{example}
Suppose $R(x) = \exp (x)$ on $(-\infty, 0]$.
Then $K_R (t) = -\log (1+t/\theta)$,
$\kappa_r (X_1) = (-\theta)^{-r} (r-1)!$,
$g(\theta) = -\theta^{-1}$,
$\sigma(\theta)^2 = -\theta^2$
and $\ell_r = (-)^r (r-1)!$ for $r>2$.
So, $\xi_n$, $\eta_n$, $g_r$, $f_r$, are independent of $\theta$.
Also $Y_n (\theta) = (\theta \overline{X}_n +1)n^{1/2}$.
So, by equation (\ref{2.20}) a confidence interval with level $1-\alpha+O(n^{-(j+1)/2})$ is
\begin{eqnarray*}
N_{nj} \left( x_2 \right) / \left |\overline{X}_n \right | \geq \theta \geq N_{nj} \left( x_1 \right) / \left |\overline{X}_n \right |,
\end{eqnarray*}
where $N_{nj} (x) = 1-n^{-1/2} \eta_{nj} (x) =1 - \sum_{i=1}^{j+1} n^{-i/2} g_{i-1} (x)$,
where $\{ g_i \}$ are given by (\ref{2.11}), (\ref{2.16}).
In this particular example, one may use
$n\theta |\overline{X}_n| \sim \Gamma (x,n)$ and hence $2n\theta |\overline{X}_n| \sim \chi^2_{2n}$
to obtain a confidence interval directly.
The expansion $L_n (x) = n+\sqrt{2} \sum_{i=1}^7 n^{-i/2} g_{i-1} (x)$ for
$\chi^2_n$ is given by equation (3a) of Fisher and Cornish (1960).
So, $\chi^2_{2n} \leq L_{2n} (-x)$ with probability $1-\Phi (x) + O(n^{-7/2})$.
It follows that for this example in terms of (3a), $N_{n6} (x) = L_{2n} (-x)/(2n)$.
\ $\Box$
\end{example}

\begin{example}
Suppose $R(x) = x^\nu$ on $[0,1]$, where $\nu >0$ is a given.
This corresponds to $h(x) = F_0 (x)^{1/\nu}$.

So, it will be both less efficient and less robust than the choice of Example 4.1.
However, it serves to illustrate the method when
$\eta_n (\cdot)$ depends on $\theta$.
In this case the cumulants are best calculated from $EX_1^r = (1+r\psi)^{-1}$,
where $\psi = (\nu \theta)^{-1}$.
So, $g(\theta) = (1+\psi)^{-1}$ lies in $[0,1]$.
Set $t=g(\theta)$.
Then
\begin{eqnarray*}
&&
\sigma (\theta)^2 = \left( 1+2\psi \right)^{-1} -t^2 = t(1-t)^2 (2-t)^{-1},
\\
&&
K_3 (\theta) = (1+3\psi)^{-1} - 3(1+2\psi)^{-1} t + 2t^3 = t(3-2t)^{-1} -3t^2(2-t)^{-1} + 2t^3,
\end{eqnarray*}
and so forth.
By (\ref{2.29})--(\ref{2.31}), $Q_1$ is given in terms of
\begin{eqnarray}
M_1 (t) = (1-t) \left( 2t^{-1}-1 \right)^{-1/2}
\label{4.6}
\end{eqnarray}
and $Q_2$ is given in terms of
\begin{eqnarray}
M_2 (t) = 2(1-t)(1-2t)/(3-2t)
\label{4.7}
\end{eqnarray}
and
\begin{eqnarray}
D_t M_1 (t)^2/2 = (2-t)^{-2} -t = \left( 2-t^{-2} \right) (1-t) \left( 1-3t+t^2 \right).
\label{4.8}
\end{eqnarray}
Finally, by (\ref{2.22}), a confidence interval of level $1-\alpha + O(n^{-1})$ is given by
\begin{eqnarray*}
\left( S_{nx_1 1} \left( \overline{X}_n \right)^{-1} - 1 \right)^{-1} / \nu < \theta < \left( S_{nx_2 1} \left( \overline{X}_n \right)^{-1} - 1 \right)^{-1} / \nu,
\end{eqnarray*}
where $S_{nx1} (t)$ is given by (\ref{2.16}), (\ref{2.23}), (\ref{2.29}), (\ref{2.30}) and (\ref{4.6})--(\ref{4.8}).
By (\ref{3.5}), the asymptotic efficiency of this choice is
$\{ V(\theta, h)$ for the maximum likelihood estimate $\} / V(\theta, h) = \theta^2 \dot{g} (\theta)^2 / \sigma (\theta)^2 = 1 - (\nu \theta + 1)^{-2}$.
\ $\Box$
\end{example}

\end{document}